\documentclass{article}

\usepackage{arxiv}

\usepackage[utf8]{inputenc} 
\usepackage[T1]{fontenc}    
\usepackage{hyperref}       
\usepackage{url}            
\usepackage{booktabs}       
\usepackage{amsfonts}       
\usepackage{nicefrac}       
\usepackage{microtype}      
\usepackage{lipsum}		
\usepackage{graphicx}
\usepackage{natbib}
\usepackage{doi}

\title{Micro-scale Electrostatic Structures formed\\ on the Rough Surfaces of the Moon}

\author{ \href{https://orcid.org/0000-0001-6491-1012}{\includegraphics[scale=0.06]{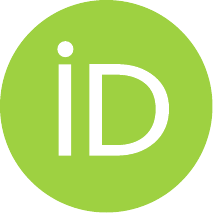}\hspace{1mm}Yohei Miyake} \\
	Graduate School of System Informatics\\
	Kobe University\\
	Kobe 657-8501, Japan \\
	\texttt{y-miyake@eagle.kobe-u.ac.jp} \\
	\And
	Jin Nakazono \\
	Graduate School of System Informatics\\
	Kobe University\\
	Kobe 657-8501, Japan \\
	\texttt{j-nakazono@stu.kobe-u.ac.jp} \\
	\And
        \href{https://orcid.org/0000-0001-7998-1240}{\includegraphics[scale=0.06]{orcid.pdf}\hspace{1mm}Yoshizumi Miyoshi} \\
	Institute for Space-Earth Environmental Research\\
	Nagoya University\\
	Nagoya 464-8601, Japan \\
	\texttt{miyoshi@isee.nagoya-u.ac.jp} \\
	\And
        \href{https://orcid.org/0000-0002-4001-6352}{\includegraphics[scale=0.06]{orcid.pdf}\hspace{1mm}Yuki Harada} \\
	Graduate School of Science\\
	Kyoto University\\
	Kyoto 606-8502, Japan \\
	\texttt{haraday@kugi.kyoto-u.ac.jp} \\
	\And
        \href{https://orcid.org/0000-0001-5992-445X}{\includegraphics[scale=0.06]{orcid.pdf}\hspace{1mm}Masaki N~Nishino} \\
	Institute of Space and Astronautical Science\\
	Japan Aerospace Exploration Agency\\
	Sagamihara 252-5210, Japan \\
	\texttt{nishino@stp.isas.jaxa.jp} \\
	\And
        Satoshi Kurita \\
	Research Institute for Sustainable Humanosphere\\
	Kyoto University\\
	Uji 611-0011, Japan \\
	\texttt{kurita.satoshi.8x@kyoto-u.ac.jp} \\
 	\And
        Satoshi Kasahara \\
	Graduate School of Science\\
	University of Tokyo\\
	Tokyo 113-0033, Japan \\
	\texttt{s.kasahara@eps.s.u-tokyo.ac.jp} \\
 	\And
	\href{https://orcid.org/0000-0001-5846-9109}{\includegraphics[scale=0.06]{orcid.pdf}\hspace{1mm}Hideyuki Usui} \\
	Graduate School of System Informatics\\
	Kobe University\\
	Kobe 657-8501, Japan \\
	\texttt{h-usui@port.kobe-u.ac.jp} \\
 	\And
        Aiko Nagamatsu \\
	Japan Aerospace Exploration Agency\\
	Tsukuba 305-8505, Japan \\
	\texttt{nagamatsu.aiko@jaxa.jp} \\
 	\And
        \href{https://orcid.org/0000-0003-4367-696X}{\includegraphics[scale=0.06]{orcid.pdf}\hspace{1mm}Satoko Nakamura} \\
	Institute for Space-Earth Environmental Research\\
	Nagoya University\\
	Nagoya 464-8601, Japan \\
}

\renewcommand{\shorttitle}{\textit{arXiv} Template}

\hypersetup{
pdftitle={A template for the arxiv style},
pdfsubject={q-bio.NC, q-bio.QM},
pdfauthor={David S.~Hippocampus, Elias D.~Striatum},
pdfkeywords={First keyword, Second keyword, More},
}

\begin{document}
\maketitle

\begin{abstract}
It is widely accepted that the surface potential of the lunar dayside is ``on average'' several to $10\ \mathrm{V}$ positive due to photoelectron emission in addition to the solar wind plasma precipitation. Recent studies, however, have shown that an insulating and rough regolith layer tends to make positive and negative charges separated and irregularly distributed on sub-Debye-length scales. The local charge separation then gives rise to an intense and structured electrostatic field. Such micro-scale electrostatic structures lie in the innermost part of the photoelectron sheath and may contribute to the mobilization of the charged dust particles. Since the electrostatic structures can take different states depending on the topography of the lunar surface, it is necessary to update the research approach. We have launched a research group to develop an integrated assessment framework that includes theoretical and numerical modeling, on-orbit observations, ground-based testing, and the development of charging measurement instruments, with the ultimate goal of comprehensively understanding the surface charging processes on the Moon.
\end{abstract}

\keywords{Lunar charging and electrostatics \and Topography effect \and Surface roughness \and Plasma accessibility \and Differential charging \and Charge separation \and Future prospects}

\section{Introduction}

Mission preparation for lunar lander exploration is rapidly increasing, and there should be a strong demand for an accurate understanding of the electrostatic environment. The lunar surface, which has neither a dense atmosphere nor a global magnetic field, gets electrically charged by the collection of surrounding charged particles from the solar wind or the Earth's magnetosphere \citep{manka1973}. As a result of the charging processes, the surface regolith particles behave as ``charged dust grains''. Dust particles have been suggested to have adverse effects on exploration instruments and living organisms during the lunar landing missions, and their safety assessment is an issue to be resolved for the realization of sustainable crewed lunar exploration \citep{collwell2007, levine2021}. It is necessary to develop a comprehensive and organized understanding of lunar charging phenomena and the electrodynamic properties of charged dust particles.\citep{garret1981}

The understanding of the lunar surface charging has been developed on the basis of similarities in terms of physical processes, with spacecraft charging and probe theory in space environments, dusty plasmas, as well as the wall-plasma interactions in laboratory plasmas \citep{whipple1981, garret1981, goertz1989, hutchinson2002, lieberman1994}. On the other hand, some studies have been focused on unique features of the surface charging on the Moon and other regolith-covered airless planetary bodies \citep{de1977, criswell1977, farrell2007, zimmerman2011, poppe2012a, zimmerman2012, miyake2015, wang2016, zimmerman2016}. The intent of this paper is, first, to highlight some unique aspects of lunar charging that remain as open questions and, second, to organize the pros and cons of the respective research approaches that are used as a method for studying surface charging phenomena. Finally, we discuss the direction for integrating different approaches to elucidate the complex electrostatic environment on the Moon.

\section{Global Picture of Lunar Surface Charging}

With no dense atmosphere or global magnetic field, the lunar surface is directly exposed to plasma charged particles from outer space. The majority of the charge carried by the plasma particles that attain the lunar surface is captured by the surface of regolith particles. It follows that the lunar surface gets electrically charged, and its surface potential differs from that of the outer space \citep{manka1973}. The finite potential difference (or simply, the surface potential) is the consequence of charge transport between the lunar surface and space, and in this aspect, it is essentially the same type of physical phenomenon as spacecraft charging. The current from space to the lunar surface depends on both the space environmental conditions and the lunar surface potential itself. The electrostatic field generated on the charged lunar surface attracts or repels surrounding plasma-charged particles. Consequently, the amount of current is a function of the lunar surface potential. The net charge transport between the environment and the surface and the associated change in the surface potential should proceed until the net current to the lunar surface reaches zero \citep{whipple1981}.

\begin{figure}[t]
	\centering
        \includegraphics[width=0.5\textwidth]{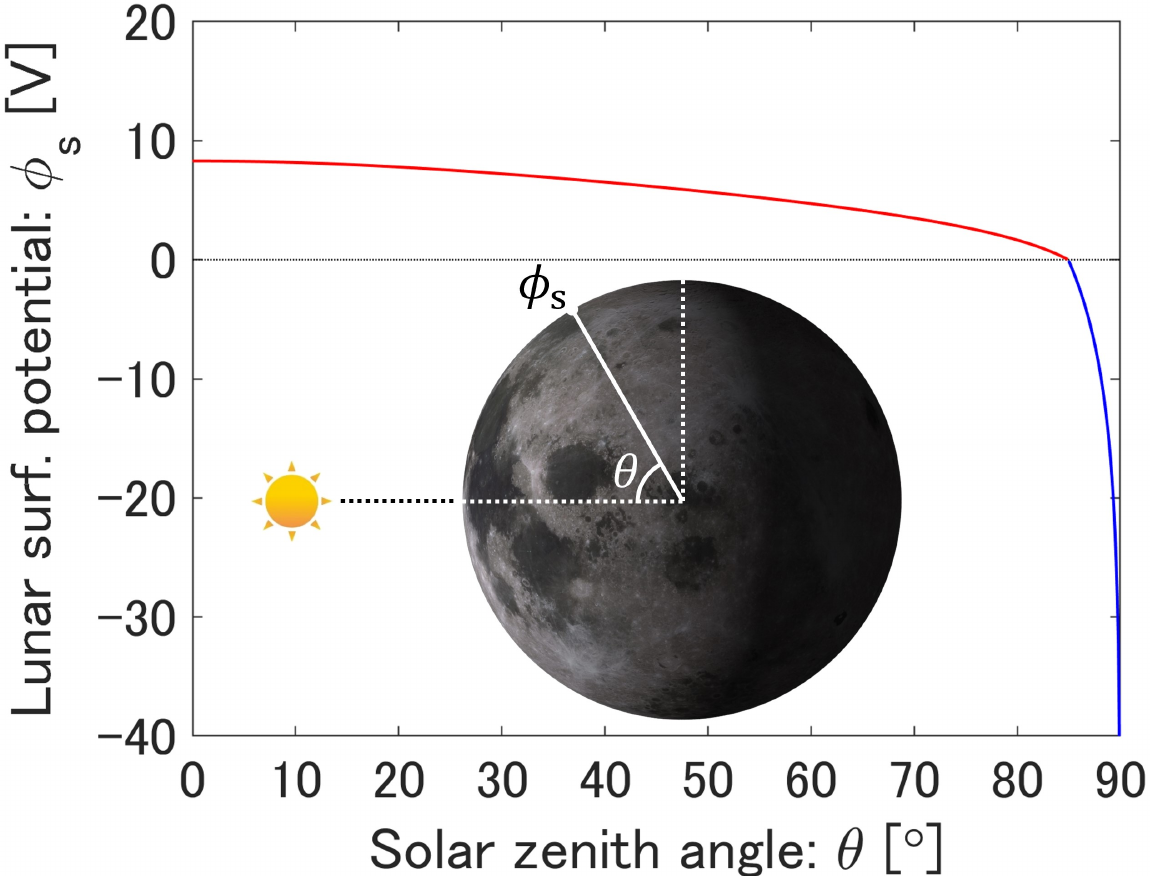}
	\caption{Dependence of the dayside lunar surface potential $\phi_\mathrm{s}$ on the solar zenith angle $\theta$, derived from the current balance condition between the photoelectron current $J_\mathrm{ph}$, the solar wind (SW) electron current $J_\mathrm{e}$, and the SW ion current $J_\mathrm{i}$ as expressed as the functions of $\phi_\mathrm{s}$. The red and blue lines correspond to the domains of $J_\mathrm{ph}$ and $J_\mathrm{e}$ dominance, respectively. It is assumed that the photoelectron temperature $T_\mathrm{ph}=2.2\ \mathrm{eV}$, the SW electron temperature $T_\mathrm{e}=10\ \mathrm{eV}$, the nominal photoelectron current density $J_\mathrm{ph0}=4.5\ \mathrm{\mu Am^{-2}}$, the SW plasma density $n_0=10\ \mathrm{cc^{-1}}$, and the SW bulk flow velocity $V_\mathrm{sw}=400\ \mathrm{kms^{-1}}$.}
	\label{fig:szadepend}
\end{figure}

In finite-temperature plasmas, the inflow electron flux is typically greater than that of ions, resulting in negative charging in situations where it is not necessary to consider electron emission processes from the surface. In contrast, the surface is positively charged in situations where photoelectron (and other electron) emission from the surface is the dominant current term. This overall picture has been actually observed for the Moon, where the day side of the moon, which is exposed to solar ultraviolet radiation, acquires a positive potential of several volts to several tens of volts \citep{freeman1973, reasoner1973, freeman1975, collier2011, halekas2011, poppe2012b, harada2017}. This positive potential is due to the photoelectron emission. It is greatest at the sub-solar point, where the solar irradiation flux is greatest, and decreases as the solar zenith angle increases. A simple estimate indicates that the plasma electron current exceeds the photoelectron current at zenith angles higher than $80$ degrees, resulting in a transition of the lunar potential to negative values, as shown in Figure~\ref{fig:szadepend}. Since the photoelectron effect does not act on the night side of the Moon, the surface potential is negative there. The night-side lunar potential in the solar wind is intimately connected to the physics of the wake caused by the plasma flow being obstructed by the Moon. This makes analytical estimation of the surface potential more difficult. (There have been cases where this has been attempted, such as the work conducted by \citet{farrel2010}). Negative potentials up to the order of -100 V have been identified by means of remote surface potential estimation from lunar orbiters \citep{freeman1975, lindeman2014, benson1977, halekas2002, halekas2008, nishino2017}.

\section{Anomalous Charging in Deep Depressions}

The Moon, a natural celestial body, exhibits a wide range of topographical features such as craters, their rims, deep pits, and boulders, resulting in diverse surface morphologies. At more local scale, the regolith accumulation layer is not flat but rather full of bumps and dips. At an even finer scale, the voids between the regolith particles appear to be deep depressions or cavities formed on the surface. From the perspective of charge transport processes, such various surface shape patterns lead to differences in the accessibility of incoming charged particles from space. The fact that the incoming charged particle flux differs from one part of the rough lunar surface to another implies that the lunar surface potential should be inhomogeneous on the spatial scale of the surface irregularities in shape. The surface of regolith particles can be considered insulating in comparison to the space filled with plasma ionized gas \citep{olhoeft1974}. It follows that the potential distribution formed on the irregular lunar surface does not easily relax or smooth out, creating a complex electrostatic field structure.

The irregularities of the local surface potential caused by craters and boulders on the Moon have been investigated through both theoretical and numerical approaches \citep{de1977, criswell1977, farrell2007, zimmerman2011, miyake2015, zimmerman2016}. The topography effects do not merely produce a change in the potential value of each part but may significantly transform a current equilibration regime itself. Even a commonly accepted plasma current magnitude ordering, i.e, $J_\mathrm{ph0} > J_\mathrm{e0} > J_\mathrm{i0}$, where $J_\mathrm{ph0}$, $J_\mathrm{e0}$, and $J_\mathrm{i0}$ represent photoelectron, background electron, and ion currents to an uncharged surface, respectively, can be disrupted by considering certain classes of surface geometry. Our recent numerical investigations have demonstrated such cases: deep depressions that can be assumed to exist in a variety of lunar surface conditions exhibit a charging state that differs greatly from the conventional surface charging state of a surface in sunlight \citep{nakazono2023}. The study, which is outlined below, focused on the differences in the directionality of collective motion of electrons and ions in the solar wind.

\begin{figure}[t]
	\centering
        \includegraphics[width=0.5\textwidth]{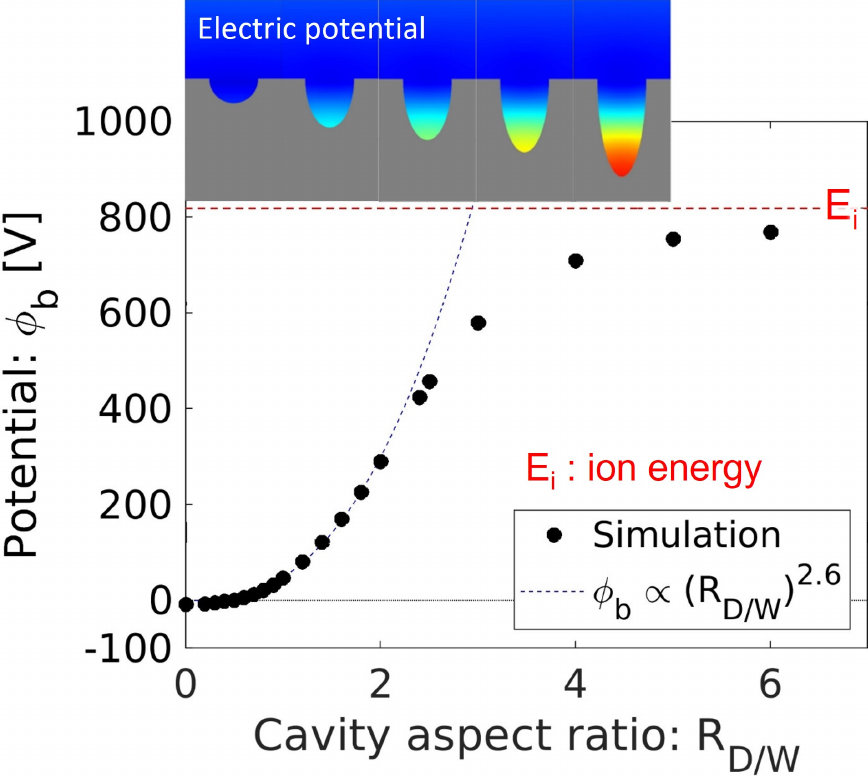}
	\caption{Model-predicted positive charging anomalously developing at the depths of deep cavities exposed to a solar wind plasma. The plot represents the dependence of the cavity bottom potential on the cavity depth-width aspect ratio, derived from a number of particle-in-cell simulation runs. Figure adapted from \citet{nakazono2023}}
	\label{fig:cdepress}
\end{figure}

The bulk velocity of the solar wind, $V_\mathrm{sw}$, and the thermal velocities of electrons and ions, $V_\mathrm{te}$ and $V_\mathrm{ti}$, satisfy the magnitude ordering $V_\mathrm{te} > V_\mathrm{sw} \gg V_\mathrm{ti}$. Therefore, the ion velocity distribution as a group can be considered to be directional (or ``beam-like''), while the electron velocity distribution is considered to be non-directional. When this solar wind plasma enters a deep depression on the lunar surface from above, electrons are preferentially captured by the sidewalls and lost near the entrance to (or the shallow part of) the depression, whereas ions are able to penetrate deep into the depression. This implies that the deeper the depression, the greater the survival rate of ions as free particles exceeds that of electrons. Consequently, at the bottom of the depression with a depth exceeding a certain threshold, the condition $J_\mathrm{i0} > J_\mathrm{e0}$ can be hold, which is not typically encountered for a flat surface in finite-temperature plasmas. Under these conditions, positive charging can be pronounced even without the involvement of electron emission processes (such as photoemission). Notably, the maximum potential in this regime reaches the order of $+\mathrm{kV}$, which is comparable to the bulk kinetic energy of the solar wind ions. The results of a numerical study indicating this are presented in Figure 2. The results demonstrate the maximum potential on the bottom surface corresponding to the depth-to-diameter aspect ratio $R_{D/W}$. A rapid increase in potential is observed when $R_{D/W}$ is $2$ or greater, and a maximum positive potential of $+800\ \mathrm{V}$ was confirmed to be achieved (Figure~\ref{fig:cdepress}).

A detailed inspection of the charged particle dynamics revealed that the physical origin of the highly-positive potential is the significant loss of free electrons before approaching the innermost part of the deep depressions, which is caused by shadowing of electrons by the shallower part of their sidewalls. This shadowing effect should be in principle governed by a dimensionless parameter, i.e., the aspect ratio, rather than the size itself of the depression. Moreover, it can be postulated that rather small, sub-Debye-scale cavities should better fit this simple physical picture, as particle trajectories are less influenced by space charges inside the depression. In light of this brief discussion, it is reasonable to assume that the ion-driven charging mechanism works effectively for micro-scale depressions or cavities. The ongoing investigation will address this aspect by examining the dependence of the charging effect on the size of depressions.

The numerical results indicate the possibility for high potentials to exist within the deep depressions. In practice, several physical factors that would mitigate the anomalous charging should be taken into account. One such factor is the emission of electrons, which can occur through photo- and secondary emission processes. While these processes are generally considered to drive positive charging, they will rather act to relieve ion-driven positive charging in this specific context. The effect is brought by an electron behavior such that electrons emitted at a certain patch of the lunar surface are recollected at another patch with a high potential. This action tends to smooth out the differential charging over the insulating surface \citep{grard1997}. It should be noted that the action of the aforementioned behavior is likely to be strongly dependent on more detailed surface geometry parameters. The generalization of its contribution is therefore left as a matter for future investigations.

\section{Spatial Hierarchy in Lunar ES Environment}

The distinctive feature of the ion-driven charging is that surface patches with positive charges are lying only in the depths of the depression. The shallow surface of the depression, in contrast, exhibits a pronounced accumulation of negative charge, as incoming electrons are preferentially collected at these locations. This indicates that the peculiar charging effect is the consequence of localized charge separation that takes place within the specific surface morphology. Such localized potential difference may lead to an intense electrostatic field, which suggests possible implications for dust grain mobilizations. We should note that the result highlighted above is considered to be just one of many forms of local charge separation effects that may produce intense electrostatic fields. It is reasonable to consider that the effect coexists with other charging forms such as proposed by \citet{wang2016}, depending on, e.g., a lighting condition of each surface patch. A critical implication from the series of studies is that insulating surfaces with irregular geometries can produce irregular potential distributions due to sub-Debye scale charge separation.

\begin{figure}[t]
	\centering
        \includegraphics[width=0.8\textwidth]{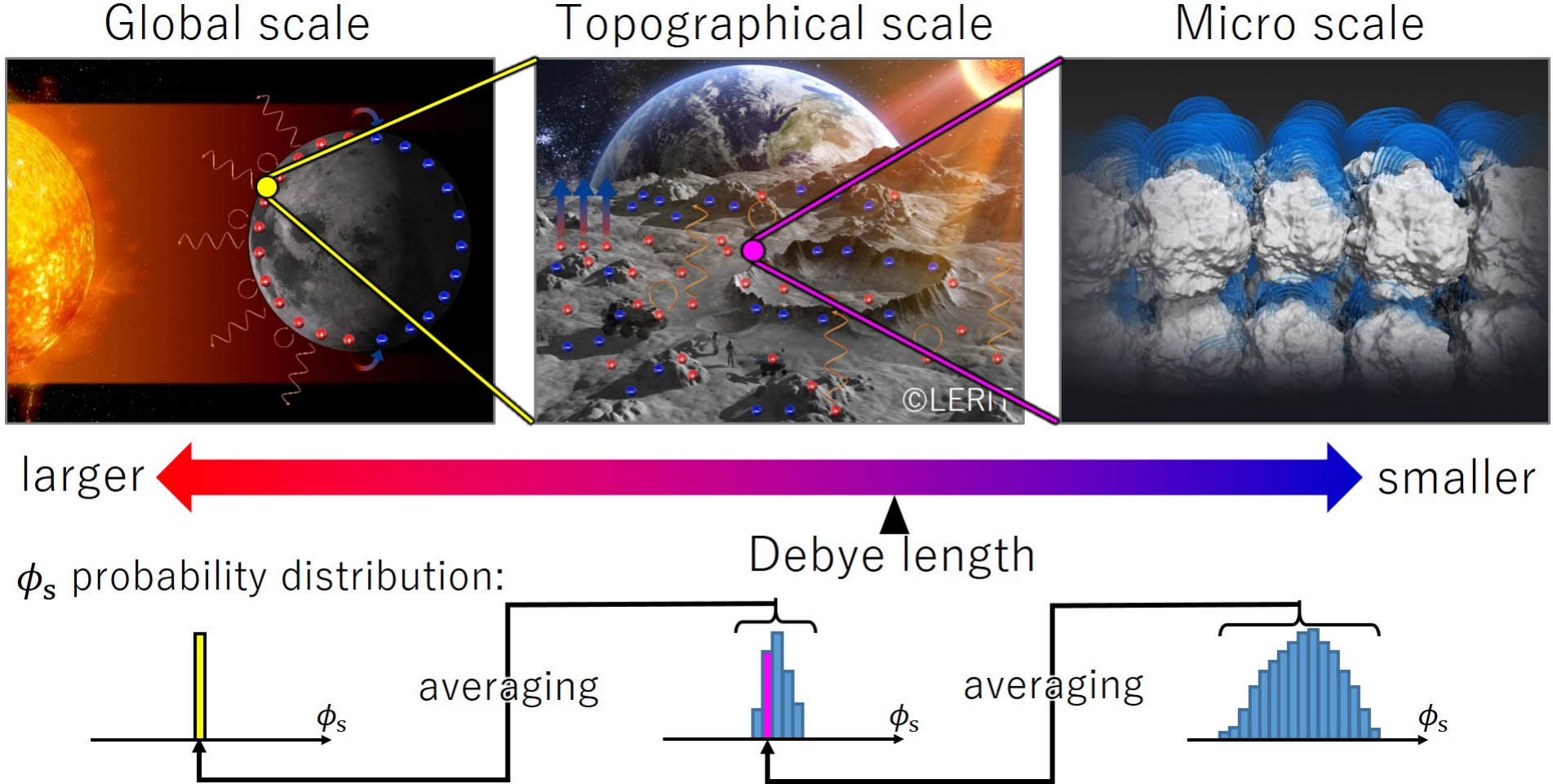}
	\caption{Concept of a hierarchical electrostatic structure anticipated on the Moon. The bottom illustration schematically shows the spread of the probability distribution of the surface potentials in each spatial scale range. The specific shape of the distribution is currently unknown and should be addressed in future studies.}
	\label{fig:eshierarchy}
\end{figure}

Given that such peculiar charging states are due to localized charge separation, it can be inferred that averaging them over space restores the well-known charging states discussed in the extensive works. Thus, it is possible to propose a certain hierarchical structure, as shown in Figure~\ref{fig:eshierarchy}, for the charging state of a celestial body covered with irregular and insulating regolith particles, such as the Moon.

\begin{description}
\item[Global scale] The day side, especially the sub-solar point, has a positive potential maximum comparable to the photoelectron energy, and the potential decreases as one approaches the day-night boundary. Then, the polarity of the charge turns negative near the solar zenith angle of $80$ degrees.

\item[Topographical scale] A closer look at a particular region will identify a meso-spatial scale differential charging caused by medium-scale topographic effects such as craters.

\item[Micro scale] A more detailed examination of the charging state over the regolith floor at a given point on a certain topography will show a more manifest variance of local potentials. Such variance should reflect the more extreme surface morphology possible at the microscale, while still being centered on the ``spatially averaged'' potential examined at the topographic scale.
\end{description}

The proposed view of the lunar electrostatics is a natural consequence of differences in the proximity of charged particles to different surface patches of the Moon or other planetary bodies with complex surface morphology, as well as differences in the conditions of sunlight exposure. On the other hand, it should also be noted that there are several effects that tend to mitigate such charge separation effects. First, as mentioned earlier, electron emission and recollection from the regolith floor can be considered. Once emitted, the electrons are recaptured at other points, forming current bridges \citep{grard1997}. These current bridges would generally act to mitigate the differential charging on the surface. The finite conductivity of the regolith itself is also important. Regarding conductivity, we mentioned earlier that it can be neglected (i.e., the lunar surface can be considered insulating). In fact, whether such an approximation holds in practice depends on the magnitude of the plasma current that can reach the point under consideration. Plasma particles precipitating from outer space (or even photoelectrons emitted outside the depression) are generally difficult to reach the depths of deep depression. The amount of their current is essentially small. In the extreme limit, there should be situations where the conductivity of the regolith surface cannot be neglected.

There is also the possibility that other physical processes could mitigate the micro-scale differential charging. For instance, differential charging could be relaxed by electrostatic breakdown and discharges that could occur between surface patches with a significant potential difference \citep{zimmerman2016}. Another possible scenario is that the intense electrostatic forces give rise to a reorganization of the stacked structure of regolith particles, as such a way that the potential difference is relaxed [Hor\'{a}nyi, private communication]. Although neither of these cases is beyond the realm of possibility at the current state of research. They are a matter that should be examined in the future as the intriguing forms of releasing processes of electrostatic energy stored near the surface of the regolith layer.

\section{Research Approaches for Lunar Electrostatics}

An integrated approach that leverages the strengths of the features of modeling, on-orbit observation, and ground-based testing approaches is crucial for comprehensively understanding the multi-scale features of the lunar charging. A significant challenge for the micro-scale electrostatic structure is the lack of observational or experimental data that provide evidence of the physical processes derived from the model predictions. A significant feature of the ion-driven charging is in the fact that the potential field itself is lying in the depths of the depressions and hidden from the outer space. Nevertheless, an intense electric field within the depressions potentially has an ability to accelerate positive charged particles upward. Thus remote observations of ascending ions or lofted dust grains and their correlations with the roughness or porosity properties in certain areas of the Moon may provide evidence for the proposed scenario.

\begin{figure}[t]
	\centering
        \includegraphics[width=0.85\textwidth]{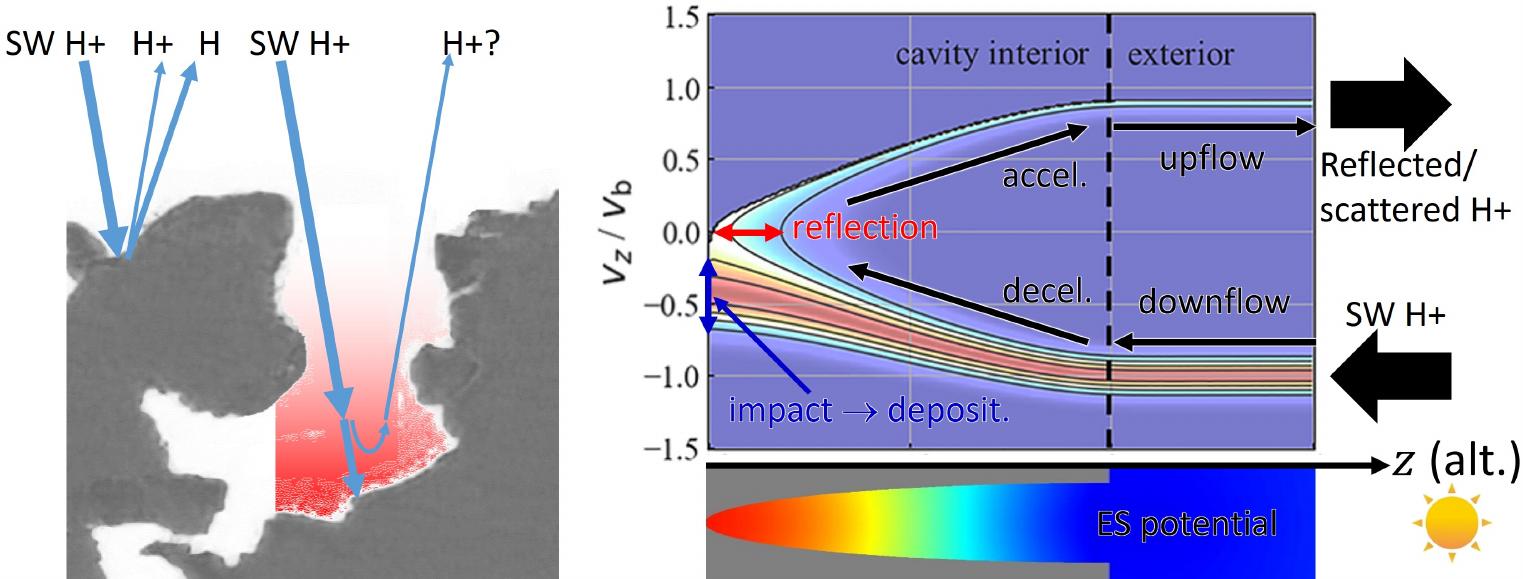}
	\caption{(Left) Extreme charging conditions that occur locally on the irregularly shaped lunar surface can cause proton reflection/scattering due to purely electrostatic action, in addition to conventional proton scattering due to direct interaction with lunar surface constituents. (Right) The phase plot of ions simulated along the vertical axis intersecting the center of the cavity bottom \citep{nakazono2023} The precipitating proton population is seen in the negative $v_z$ domain. Inside the cavity ($z<0$), the deceleration of the descending flow gets pronounced, which is the consequence of the intense upward electric field. Despite the deceleration, the majority of the ions reach the cavity bottom surface whereby they contribute to the positive charging. Meanwhile, a small portion of the ions are electrostatically reflected before reaching the bottom and form an ion upflow.}
	\label{fig:upflow}
\end{figure}

Some results are beginning to come out that can serve as a starting point for such discussions. It is a well-known observational fact that a small fraction of the solar wind protons precipitating on the lunar surface are scattered while retaining their charged state \citep{saito2008, lue2018}. Most previous studies have attempted to explain this phenomenon in terms of plasma-material interactions at the time of direct proton impact on lunar surface constituents. Our numerical results show that the localized anomalous charging also raises the possibility that some ions are reflected from a positive potential barrier before impacting the surface (Figure~\ref{fig:upflow}). This is a purely electrostatic reflection and scattering process. It will be a challenging task to demonstrate that such electrostatic reflection/scattering contributes in part to the observed ion scattering. One approach would be to see if there is a correlation between the observations and the model prediction of micro-scale charging.

To date, the ground-based testing may be the most adapted and direct method for detailing the micro-scale differential charging associated with the irregularities of the regolith floor. Vacuum chamber experiments have been conducted, in which a plasma flow or UV light have been irradiated on the bed of a lunar soil simulant \citep{wang2016, yu2015, hood2018, orger2019}. This enables the observation of electrostatically driven dust mobilizations in a controlled environment. On the other hand, what is generally obtained from such ground testing is not direct information about the charging states on individual regolith particles, but some information about the mechanical events induced by the electrostatic energy stored in the aggregate. A cross comparison between the numerical simulations and the ground testing will provide an opportunity to validate the numerical model as well as physical insight on the micro-scale charging and the associated dust mobilizations.

Further innovations in numerical modeling itself are also required. Numerical analysis based on the commonly used particle-in-cell model is not necessarily efficient in solving the charging process of sub-Debye scales. It is necessary to develop a dedicated approximate model that can shorten the analysis time and put it into practical use. Such an approach could introduce measures to emulate and speed up numerical analysis with the aid of machine-learning-based surrogate models. Additionally, the investigations should address the numerous degrees of freedom regarding the surface structural features of the Moon and other airless planetary bodies, not only deterministic simulations as addressed herein but also some statistical approaches would be necessitated.

\section{Conclusions}

The physics of lunar charging began with an examination of the electrostatic structure on a global scale. On this scale, the Moon can be assumed to have a smooth surface, and the knowledge accumulated from extensive research in probe theory and spacecraft charging can be applied in a straightforward manner. As research in this field has progressed, more attention has been paid to detailed electrostatic structures on smaller spatial scales. This trend will become even more pronounced with the recent acceleration of lunar exploration, including crewed landing missions. Our contention in this paper is that the electrostatic environment near the lunar surface will be characterized by localized deviations from the spatially averaged potential. This perspective would be important because the potential difference in small distances is directly associated with an intense electrostatic field, and because human activities on the Moon is about to take place in just such a complex electrostatic environment. Understanding the micro-scale electrostatic environment is essential for assessing the variety of risks that may be encountered during human activities on the Moon and for developing methods to mitigate them. Independent investigations using only a single research approach alone are limited in addressing this challenging issue. The establishment of an integrated assessment framework that organically combines on-orbit observations, ground testing, and numerical modeling is truly needed.

\section*{Acknowledgments}
The present study was supported in part by the Japan Society for the Promotion of Science: JSPS (Grant No. 20K04041), the innovative High-Performance Computing Infrastructure: HPCI (Project No. hp230015 and hp240065), and the Joint Usage/Research Center for Interdisciplinary Large-scale Information Infrastructures: JHPCN (Project No. jh240016) in Japan.
Y. Miyake thanks Mihaly Hor\'{a}nyi, Xu Wang, and Jan Deca for fruitful discussions on the mechanisms of lunar charging and dust mobilization.
Y. Miyake would like to extend his deepest appreciation to Hiroshi Nakashima, who passed away in 2021, for his long-term support on the use of state-of-the-art HPC technologies.

\bibliographystyle{unsrtnat}
\bibliography{manuscript}  






\end{document}